\documentstyle[12pt,qqa4scsk]{article}
\input tcilatex

\QQQ{Language}{
American English
}

\begin{document}

\author{P. N. Bhat, P.R. Vishwanath, B. S. Acharya \\ 
and M. R. Krishnaswamy   \and {\it Tata Institute of Fundamental Research}\\%
{\it Homi Bhabha Road, Mumbai 400 005, India.}}
\title{PACHMARHI ARRAY OF \v CERENKOV TELESCOPES}
\maketitle

\begin{abstract}
Very High Energy (VHE) $\gamma -$ray Astronomy observations are planned to be
carried out at Pachmarhi in the central Indian state of Madhya Pradesh using
the well known atmospheric \v Cerenkov technique. An array of 25 \v Cerenkov
telescopes is currently under construction. Using this array it is proposed
to sample the \v Cerenkov light pool at various distances from the shower
core in order to estimate the lateral distribution parameters of the shower.
In this paper we discuss the scientific motivation of this concept of
enriching the $\gamma -$ray signal as compared to the standard imaging
technique. The current status of the detector development and the expected
results will be presented.
\end{abstract}

\section{INTRODUCTION}

The $\gamma -$ray region of the electromagnetic spectrum has been the last
to be successfully exploited as a channel for astronomical investigation.
Although it was realized as far back as 1958 (Morrison, 1958), the detection
techniques were difficult and the fluxes were low. The chief motivation for
the study of $\gamma $-rays is that they are produced in the high energy
particle interactions. The close association between cosmic rays and $\gamma 
$-rays makes $\gamma $-ray astronomy a special branch of cosmic ray
astrophysics. A careful study of the intensity spectrum of $\gamma $-rays
and their time variability may reveal the physical conditions and
acceleration mechanisms of cosmic rays with in the source.

VHE $\gamma -$ray astronomy, covering the energy range 10$^2$-10$^4$ GeV,
has generated a lot of interest as many new groups all over the world are
entering the field. Further the field has come of age and now one can say with
certainty that some VHE $\gamma $-ray sources do exist.

\section{PACHMARHI ARRAY OF \v CERENKOV TELESCOPES}

PACT is situated at Pachmarhi ( longitude: $78^{\circ }~26^{\prime }~E$,
latitude: $22^{\circ }~28^{\prime }~N$ \& altitude: 1075 m). In a constant
effort to improve the signal to noise ratio, it was realized that one has to
reduce the background due to hadronic cosmic ray showers. Monte-Carlo
simulations have shown that the differences between hadron induced and $\gamma $%
-ray initiated cascades could be exploited to reject the former
(Hillas and Patterson 1990, Rao and Sinha 1988). The simulations
have also shown that electromagnetic cascades are flatter in the lateral
distribution of \v Cerenkov photons (Bhat, 1997) than those initiated by
cosmic ray primaries.

Several groups have used some of the above parameters to reject the cosmic
ray background and enhance the signal to noise ratio (Weekes 1989, Baillon 
et al.,1994 and T\"umer et al.,1985). The Whipple group has
detected steady emission of TeV $\gamma $-rays from Crab nebula using \v
Cerenkov imaging technique (Vacanti et al., 1991 and Punch et al.,1992) and
have established this source as a `standard candle' of TeV $\gamma -$rays.

However, not enough attention has been given to the lateral distribution
aspect of the atmospheric \v Cerenkov radiation to reject cosmic ray
initiated background. Angular imaging and spatial multiple sampling are two
complimentary ways to examine the same bundle. However one looks at
different types of distinguishing features between $\gamma $-ray and cosmic
ray primaries in the two techniques. The lateral distribution of \v Cerenkov
photons have a ``hump'' at distances of 120-140 m from the shower core in $%
\gamma $-ray initiated cascades only (Rao and Sinha 1988; Bhat, 1997).
Further, the shower to shower fluctuations in the lateral distribution of \v
Cerenkov photons is much less for a $\gamma $-ray initiated shower compared
to cosmic ray initiated showers, which are rather `bumpy' due to the
contribution of \v Cerenkov light from muons. The average lateral
distribution of \v Cerenkov photon densities clearly show a flat
distribution up to the ``hump'' region for $\gamma $-ray initiated showers
and a steeper distribution for proton initiated ones( Bhat, 1997). Rao and
Sinha (1988) have shown that the signal to noise ratio could be
improved atleast by a factor of 2, by exploiting these very differences.

\section{THE ARRAY}

We have decided to pursue the latter technique viz. to measure the lateral
distribution parameters. The entire available area of $85~m~\times ~100~m$
at the High Energy Gamma Ray Observatory (HEGRO), Pachmarhi has been filled with an array of 25 \v Cerenkov
telescopes. Each telescope consists of 7 parabolic mirrors (f/d $\sim $ 1)
mounted symmetrically making up a total reflector area of $4.4~m^2$. These
parabolic reflectors, fabricated locally, have a point source image of $<$ 1$%
^{\circ }$. The 7 reflectors are mounted on a single equatorial mount which is
steerable both in E-W and N-S directions. The reflector orientation and
tracking are controlled by a computer automated system (Gothe, et al.,
1997) to an accuracy of $\pm 0^{\circ }.1$. A fast photomultiplier (EMI
9807B) is mounted at the focal plane of each reflector behind a $3^{\circ }$
diameter mask.

The existing large and small mirrors will be deployed in the form of a
compact array at the center. These will generate an independent trigger and
hence will work like a compact array within the larger array, having a
significantly lower energy threshold. From this one will be able to derive
an energy spectrum of $\gamma -$rays from the source

Event arrival times are derived from a Global Position Satellite (GPS)
receiver having an absolute time keeping accuracy of $\pm 100~ns$. The pulse
heights and the times of arrival of pulses at each PMT will be recorded
using LeCroy ADC and TDC modules respectively. A new distributed data
acquisition system based on several PC 486's networked together using a
LINUX system is currently being developed (Bhat, 1996). Event triggers will
be generated by each of the 25 telescopes and tagged. The idea of a
distributed data acquisition is to minimize the loss of information due to
transmission through long coaxial cables. Special cables (RG 213) of shorter
lengths will be used for analog pulse transmission from the phototubes to
the nearest signal processing center. Event triggers will be generated from
the royal sum of all the analog pulses from all the 7 mirrors in a
telescope. The measurement of time differences of triggers from various
telescopes will enable us to measure the arrival direction correct to 0.2-0.3%
$^{\circ }$.

Each event data will consist of the event arrival time to an accuracy of $%
\pm 1~{\mu }s$, amplitude and relative time of arrival of pulses at each
bank, trigger information and other relevant house-keeping informations etc
will be recorded at each of the 4 field signal processing centers(FSPC) while the
relative arrival times between the telescopes will be recorded at the
central data acquisition system. The data will be recorded in the internal
hard disks which will be collated off-line.

A Computer Automated Rate Adjustment and Monitoring System (CARAMS) is
developed for setting the high voltages on individual phototubes such that a
preset rate of pulses above the set threshold of 30 mV is generated (Bhat,
1996a). This will also monitor the counting rates from all the phototubes
through out the run. Similarly, the optimum count rate from each photo tube
is estimated using the count rate variation as a function of the
discriminator threshold. A software package (Automated Rate Measurement At
different Discriminator thresholds, ARMADA) to derive this curve is also
developed. This will run prior to CARAMS before starting an observation.

\section{PHOTON \& ENERGY THRESHOLDS}

If we use a the royal sum of the analog pulses from the 7 mirrors in a
telescope to generate a trigger, it has been estimated that the trigger rate
will be around 120/min. The night sky background flux at Pachmarhi was
measured to be 1.2 $\times $ $10^8$ photons cm$^{-2}$ s$^{-1}$ (Bhat and
Mehta, 1996). Using this it is estimated that the threshold photon density
is $\sim 32\ \gamma $ m$^{-2}$. Using the estimated photon densities through
monte carlo simulations at various $\gamma $-ray energies (Ong, 1995), this
photon density threshold translates to an energy threshold of $\sim $ 350
GeV for $\gamma $-rays.

\section{TECHNICAL PROGRESS}

Eighteen of the 25 telescopes are already in position and the rest will be
erected in about 2 month's time.

The new distributed data recording system as well as networking between PC's
carrying out signal processing in the FSPC's
are being developed. Each of the FSPC's will have independent clocks which
will be synchronized with the GPS through the network. The FSPC's are
constantly monitored and controlled from the central station. Also the data
recording system in the central station communicates with the PC's which
control the high voltages on the individual phototubes as well as the PC
which controls and monitors the movement of the telescopes.

We expect to see the first light around October, 1997.

\end{document}